\documentclass[twocolumn,prl,showpacs,superscriptaddress]{revtex4}

\usepackage{amsfonts}
\usepackage[T1]{fontenc}
\usepackage[latin9]{inputenc}
\usepackage{mathrsfs}
\usepackage{textcomp}
\usepackage{latexsym}
\usepackage{amsthm}
\usepackage{amssymb}
\usepackage{amsmath}
\usepackage[all]{xy}
\usepackage{graphicx}
\usepackage{epstopdf}
\usepackage{dcolumn}
\usepackage{bm}
\usepackage{color}
\usepackage{hyperref}
\usepackage{float}
\usepackage{color}
\usepackage{textcomp}
\usepackage{amsmath}
\usepackage{amssymb}
\usepackage{hyperref}
\usepackage{float}
\usepackage{amsfonts}
\usepackage{}
\usepackage{mathrsfs}
\usepackage{textcomp}
\usepackage{latexsym}
\usepackage{amsthm}
\usepackage{amssymb}
\usepackage{amsmath}
\usepackage[all]{xy}
\usepackage{graphicx}
\usepackage{epstopdf}

\SelectTips{eu}{11}

\definecolor{yellow}{rgb}{1,1,0}
\definecolor{pink}{rgb}{0.68,0.92,1}
\definecolor{lightgreen}{rgb}{1,0.6,0.78}

\begin{document}

\title{Hybrid magic state distillation for universal fault-tolerant quantum computation}

\author{Wenqiang Zheng}
\thanks{These authors contributed equally to this work.}
\affiliation{Hefei National Laboratory for Physical Sciences at Microscale and Department of Modern Physics, University of Science and Technology of China, Hefei, Anhui 230026, China}

\author{Yafei Yu}
\thanks{These authors contributed equally to this work.}
\affiliation{Laboratory of Nanophotonic Functional Materials and Devices, LQIT $\&$ SIPSE, South China Normal University, Guangzhou 510006, China}

\author{Jian Pan}
\affiliation{Hefei National Laboratory for Physical Sciences at Microscale and Department of Modern Physics, University of Science and Technology of China, Hefei, Anhui 230026, China}

\author{Jingfu Zhang}
\affiliation{Fakult\"{a}t Physik, Technische Universit\"{a}t Dortmund, 44221 Dortmund, Germany}

\author{Jun Li}
\affiliation{Hefei National Laboratory for Physical Sciences at Microscale and Department of Modern Physics, University of Science and Technology of China, Hefei, Anhui 230026, China}

\author{Zhaokai Li}
\affiliation{Hefei National Laboratory for Physical Sciences at Microscale and Department of Modern Physics, University of Science and Technology of China, Hefei, Anhui 230026, China}

\author{Dieter Suter}
\affiliation{Fakult\"{a}t Physik, Technische Universit\"{a}t Dortmund, 44221 Dortmund, Germany}

\author{Xianyi Zhou}
\email{zhouxy@ustc.edu.cn}
\affiliation{Hefei National Laboratory for Physical Sciences at Microscale and Department of Modern Physics, University of Science and Technology of China, Hefei, Anhui 230026, China}

\author{Xinhua Peng}
\email{xhpeng@ustc.edu.cn}
\affiliation{Hefei National Laboratory for Physical Sciences at Microscale and Department of Modern Physics, University of Science and Technology of China, Hefei, Anhui 230026, China}
\affiliation{Synergetic Innovation Center of Quantum Information $\&$ Quantum Physics,
University of Science and Technology of China, Hefei, Anhui 230026, China}

\author{Jiangfeng Du}
\email{djf@ustc.edu.cn}
\affiliation{Hefei National Laboratory for Physical Sciences at Microscale and Department of Modern Physics, University of Science and Technology of China, Hefei, Anhui 230026,  China}
\affiliation{Synergetic Innovation Center of Quantum Information $\&$ Quantum Physics,
University of Science and Technology of China, Hefei, Anhui 230026, China}

\begin{abstract}
A set of stabilizer operations augmented by some special initial
states known as ``magic states", gives the possibility of universal
fault-tolerant quantum computation. However, magic state preparation
inevitably involves nonideal operations that introduce
noise. The most common method to eliminate the noise is
magic state distillation (MSD) by stabilizer operations. Here we
propose a hybrid MSD protocol by connecting a four-qubit $H$-type
MSD with a five-qubit $T$-type MSD, in order to overcome some disadvantages of
the previous MSD protocols. The hybrid MSD protocol further
integrates distillable ranges of different existing MSD protocols and extends the $T$-type distillable range to the stabilizer octahedron edges.
And it provides considerable improvement in qubit cost for almost all of the distillable range.
Moreover, we experimentally demonstrate the four-qubit $H$-type MSD protocol
using nuclear magnetic resonance technology, together with the previous five-qubit MSD experiment, to show the feasibility of the hybrid MSD protocol.
\end{abstract}


\maketitle

Decoherence and control errors are some of the major obstacles for the implementation of
scalable quantum information processing. To overcome these
obstacles, quantum fault-tolerance theory has been developed
\cite{knill98,shor}, in which the information is encoded in a
subspace of a larger Hilbert space. The subspace is fixed by a
subgroup of  the Pauli group, consisting of some Hermitian tensor
products of Pauli operators which are defined as the stabilizer of
the subspace. Logical operations are \textit{transversally}
performed on the encoded information \cite{Steane98,Knill05}, aiming
to prevent the propagation of errors within the codeblock and
further avoid correlated errors in the course of quantum error
correction. Unfortunately, only a limited set of operations, known
as \textit{stabilizer operations} (consisting of Clifford group
unitaries \cite{Gottesman98}, preparation of $\left| 0 \right\rangle
$ and measurement in the computational basis) , can be implemented in
such a fault-tolerant manner, which can not
provide a universal quantum computation according to the Gottesman-Knill theorem \cite%
{Gottesman97,Gottesman04}. This dilemma can be solved by introducing a nonstabilizer state
(not eigenstates of Pauli operators) as an ancilla, and then implementing a gate
outside the Clifford group through gate teleportation \cite{knill98}.

Preparation of a nonstabilizer state would inevitably involve
non-stabilizer operations \cite{bravyi06,howard12}, which are not
fault-tolerant and induce noise to the nonstabilizer state. The most common method
for reducing noise is to distill
noisy copies of these resource nonstabilizer states to an almost
pure nonstabilizer state with only stabilizer operations
\cite{kitaevi05,bravyi06,reichard05}. The pure nonstabilizer states
that can be prepared through distillation with only stabilizer
operations are called \textit{magic states} and the
fault-tolerant distillation for magic states is called
\textit{magic-state distillation} (MSD). So far, there are two types
of states  found to be ``magic", and they are called $T$-type and $H$-type magic states \cite{kitaevi05}.  Consequently, MSD
enables universal fault-tolerant quantum computing, and it also
opens a framework to observe what kind of quantum states can provide
universal fault-tolerant computational power
\cite{browne10,howard09,howard14}. However MSD puts a big challenge
to quantum computation as it will consume up a majority of qubit
resource in architectures \cite{Jones2012}. Much effort has been
devoted to develop economical methods to get pure magic states by concatenating two MSD protocols of the same type magic state ($H$-type)
\cite{braviy12}, and to build up effective instructions to compile
magic states and the stabilizer operations for implementing
non-Clifford operations \cite{compile1}. Besides, various MSD protocols suffer from
different shortcomings. For example, for five-qubit $T$-type MSD
\cite{kitaevi05} there is a gap between the distillable unstabilizer
states and stabilizer states, for seven-qubit $H$-type MSD
\cite{reichard05}, the polarization of the output state increases
only polynomially in the number of noisy copies at the range of
high polarization, while four-qubit $H$-type MSD \cite{reichard09}
cannot yield a nearly pure magic state.

Here we propose a novel MSD protocol by hybridizing one $H$-type MSD protocol \cite{reichard09} with one $T$-type MSD protocol \cite{kitaevi05}. This hybrid one does not only overcome the
shortcomings in the previous MSD protocols mentioned above, but also brings two additional advantages: the integration of distillable ranges of previous individual MSD
protocols and great reduction of qubit resource consumption.
Moreover, up to now, the only experiment on the five-qubit $T$-type
MSD has been implemented in nuclear magnetic resonance (NMR) system
\cite{laflamme11}. Here we report an experimental demonstration for
the four-qubit MSD by NMR to show the feasibility of this
hybrid protocol.

An arbitrary one-qubit state can be represented as
$\rho  = {{(\mathbf{1} + x{\sigma _x} + y{\sigma _y} + z{\sigma _z})} \mathord{\left/
 {\vphantom {{(I + x{\sigma _x} + y{\sigma _y} + z{\sigma _z})} 2}} \right.
 \kern-\nulldelimiterspace} 2}$,
where ${\sigma _x}$, ${\sigma _y}$, ${\sigma _z}$, and $\mathbf{1}$
denote the Pauli matrices and identity operator, $\vec{a} = (x, y,
z)$ is a dimensionless vector of length $\le 1$ that specifies the
position of the state in the Bloch sphere. One
single-qubit state $\rho$ with $\left| x \right| + \left| y \right|
+ \left| z \right| \le 1$ (forming a stabilizer octahedron
$\mathcal{O}_\text{s}$) cannot be distilled to nonstabilizer states
with only Clifford operations \cite{Browne09}. Prior MSD protocols
show that some states outside $\mathcal{O}_s$ can be distilled
towards eigenstates of Clifford gates, such as the Hadamard $H$
gate and the $T$ gate \cite{howard09,Howard11}. These eigenstates
are magic states: $H$-type with $\vec{a}_H = (0, \pm \frac{1}{\sqrt
2}, \pm  \frac{1}{\sqrt 2}), (\pm  \frac{1}{\sqrt 2}, 0, \pm
\frac{1}{\sqrt 2}), (\pm  \frac{1}{\sqrt 2}, \pm  \frac{1}{\sqrt 2},
0)$ and $T$-type with $\vec{a}_T = ( \pm \frac{1}{\sqrt 3}, \pm
\frac{1}{\sqrt 3}, \pm \frac{1}{\sqrt 3})$. Without loss of
generality, here we focus on two of them$\colon$
\begin{equation}
\begin{split}
&\text{$H$-type:}~~~{\rho _\text{H}} = \left[ \mathbf{1} + ({\sigma _x} + {\sigma _z}) /\sqrt 2 \right]/2\\
&\text{$T$-type:}~~~{\rho _\text{T}} = \left[ \mathbf{1} + ({\sigma _x} + {\sigma _y} + {\sigma _z}) /\sqrt 3 \right]/2.
\end{split}
\end{equation}
%\begin{align}
%&\text{$H$-type:}~~~{\rho _\text{H}} = \left[ \mathbf{1} + ({\sigma _x} + {\sigma _z}) /\sqrt 2 \right]/2  \nonumber \\
%&\text{$T$-type:}~~~{\rho _\text{T}} = \left[ \mathbf{1} + ({\sigma _x} + {\sigma _y} + {\sigma _z}) /\sqrt 3 \right]/2.
%\end{align}
The polarization of an arbitrary state $\rho$ in the direction of
the magic states ($H$-direction or $T$-direction) is defined as
\begin{equation}
\begin{split}
&\text{$H$-type:}~{p_\text{H}(\rho)} = 2\text{Tr}(\rho  \cdot {\rho _\text{H}}) - 1 = \left( {x + z} \right)/\sqrt 2 \\
&\text{$T$-type:}~{p_\text{T}(\rho)} = 2\text{Tr}(\rho  \cdot {\rho _\text{T}}) - 1 = \left( {x + y + z} \right)/\sqrt 3,
\end{split}
\end{equation}
%\begin{align}
%&\text{$H$-type:}~{p_{H}(\rho)} = 2\text{Tr}(\rho  \cdot {\rho _H}) - 1 = \left( {x + z} \right)/\sqrt 2 \nonumber\\
%&\text{$T$-type:}~{p_{T}(\rho)} = 2\text{Tr}(\rho  \cdot {\rho _T}) - 1 = \left( {x + y + z} \right)/\sqrt 3,
%\end{align}
which quantifies how the state $\rho$ is close to the magic states.
Given a resource of these pure ``magic" states, one can implement gates outside the Clifford group  (i.e., the $\pi /12$ phase gate for the $H$-type one and the $\pi /8$ phase gate for the $T$-type one) to enable universal quantum computation.

\begin{figure}
\centering \includegraphics[width=0.48\textwidth]{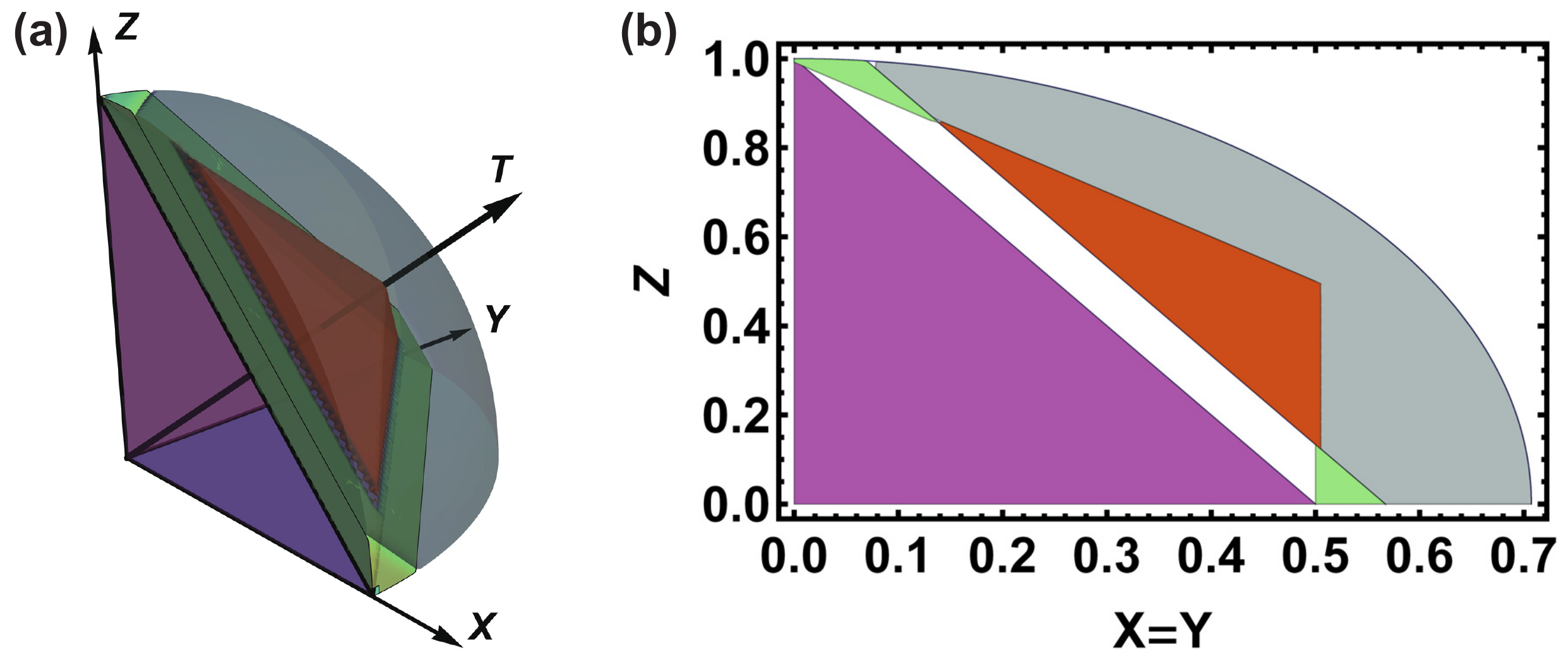} \caption{\label{fig:range}
Distillable ranges of the five-qubit $T$-type MSD protocol ($\mathcal{A}_\text{T}$, denoted as the gray and orange regions) and the seven-qubit $H$-type MSD protocol ($\mathcal{A}_\text{H}$, denoted as the green and orange regions). Purple region shows the interior of  the stabilizer octahedron $\mathcal{O}_\text{s}$.
(a) One octant of the Bloch sphere.
(b) A cross-section of one octant of the Bloch sphere through the plane $x = y$.
 }
\end{figure}

Bravyi and Kitaev proposed a $T$-type MSD protocol based on
five-qubit error correcting code \cite{kitaevi05}. Provided noisy
copies have an initial polarization in $T$-direction
$p_\text{T}(\rho_\text{in}) > \sqrt{3/7} \approx 0.655$, this
protocol yields a higher polarization. By the iteration, it is
possible to obtain the output with $p_\text{T}(\rho_\text{out}) \to
1$. We define the distillable range by this $T$-type MSD
protocol as $\mathcal{A}_\text{T}$, represented by the gray and
orange regions in Fig. \ref{fig:range}. There exhibits a gap between
the region $\mathcal{A}_\text{T}$ and the stabilizer octahedron
$\mathcal{O}_\text{s}$. In contrast, Reichardt proposed a $H$-type
MSD protocol based on the seven-qubit Steane code
\cite{reichard05,reichard09}. It is possible to obtain the output
with $p_\text{H}(\rho_\text{out}) \to 1$ by this protocol, provided
noisy copies have an initial polarization in $H$-direction
$p_\text{H}(\rho_\text{in}) >1 / \sqrt{2} \approx 0.707$. We define
this distillable range of the seven-qubit $H$-type MSD protocol as
$\mathcal{A}_\text{H}$, represented by the green and orange regions
in Fig. \ref{fig:range}. The range $\mathcal{A}_\text{H}$ is tight
(no gap) in the directions crossing the octahedron edges, which means a
transition from universal quantum computation to classical efficient
simulation \cite{howard09}. Alternatively, states can be distilled
in the $H$-direction by using a four-qubit Clifford circuit
\cite{reichard09}, at the price of a smaller distillable range
$\mathcal{A}^{'}_\text{H} \subset \mathcal{A}_\text{H}$, where the
ultimate polarization $p_\text{H}(\rho_\text{out})$ is approximately
equal to 0.964, not 1. From Fig. \ref{fig:range}, we can see that
the distillable ranges of the $H$-type and $T$-type MSD protocols do not
overlap completely. It should be observed that $
\mathcal{A}_\text{H} - (\mathcal{A}_\text{T} \cap
\mathcal{A}_\text{H}) = \mathcal{A}^{'}_\text{H} -
(\mathcal{A}_\text{T} \cap \mathcal{A}^{'}_\text{H})$.

\begin{figure}
\centering \includegraphics[width=0.49\textwidth]{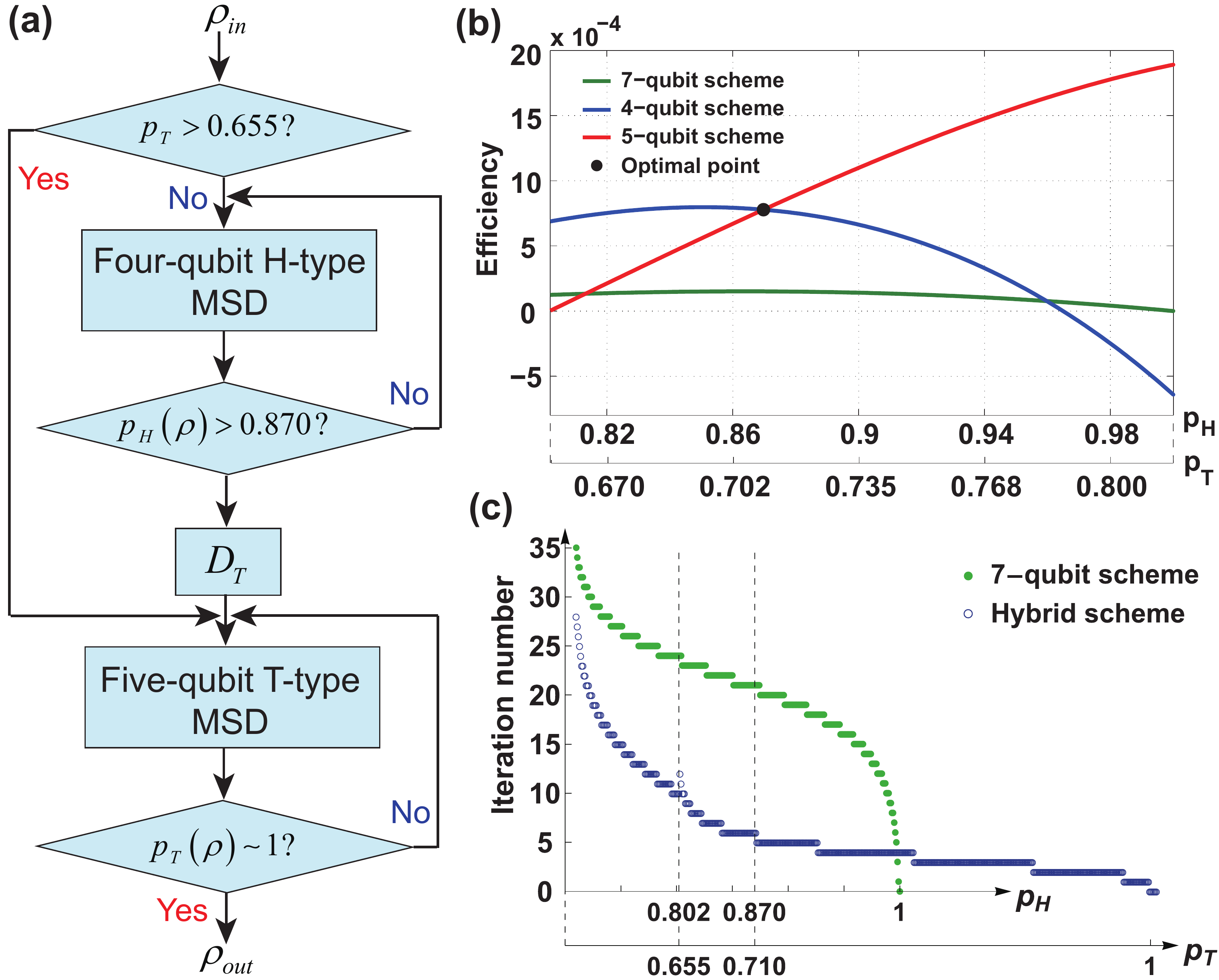} \caption{\label{fig:scheme}
(a) Flowchart of the hybrid MSD protocol ${D_\text{T}}\left( \rho  \right){\rm{ = }}\left( {\rho  + T\rho {T^\dag } + {T^\dag }\rho T} \right)/3$ is the $T$-projection operation. (b) The efficiency $\nu$ of the MSD protocols for the state lying in $H$-direction. The second horizontal axis shows the corresponding $p_\text{T}$.
The polarization loss resulted by ${D_\text{T}}$ has been considered when we calculated the efficiency of the four-qubit protocol.
(c) The necessary iteration number to distill noisy states for an almost pure magic state ($T$-type or $H$-type) with the target fidelity above 0.999.
}
\end{figure}

We found that $\mathcal{A}^{'}_\text{H} \cap \mathcal{A}_\text{T}
\ne \emptyset$ and the state with $p_\text{H}(\rho_\text{in}) >0.802$ can  surely be distilled by the $T$-type MSD protocol.
Based on this fact, we propose a hybrid protocol, whose flowchart is shown in Fig.
\ref{fig:scheme}(a). Got an ensemble of noisy magic state with the polarization
$p_\text{H}(\rho_\text{in}) > 0.707$,
%Here we consider a hybrid MSD protocol
%to extend the distillable
%range for $T$-type magic states to the wider region
%$\mathcal{A}_\text{T} \cup \mathcal{A}_\text{H}$. Figure
%\ref{fig:scheme}(a) shows the flowchart of this hybrid MSD protocol.
%Based on the ability to reprepare the noisy nonstabilizer state
%$\rho_\text{in}$,
we first choose some samples and measure their polarizations in
$T$-direction to check whether
$p_\text{T}(\rho_\text{in})>0.655$. If yes, these noisy copies are
directly distilled by the five-qubit $T$-type MSD module.
Otherwise we send them into the four-qubit $H$-type MSD
module for the higher polarization
$p_\text{H}(\rho_\text{int})$.

The distillation protocols require measurements of the code's stabilizers. Only when all measurement outcomes are ``$0$", this round of distillation is successful.
Assuming $\theta$ is the success probability, $n/\theta$ is the average qubit consumption in each iteration, where $n=4,5,7$ for the four-qubit, five-qubit and seven-qubit MSD protocols, respectively. Further, $\nu = \Delta {p}\theta /n$ is the increased polarization per consumed qubit in each iteration, representing  the efficiency of the protocol, where $\Delta {p}$ is the increased polarization in the target direction after one iteration.
%Comparing the efficiency of the three MSD protocols in Fig.\ref{fig:scheme}(b), we can see that  seven-qubit protocol  performs much lower in efficiency and in the range $0.707 < p_H < 0.870$, the four-qubit protocol is more efficient, while beyond this range the five-qubit protocol has higher efficiency.
Comparing the efficiency of the four-qubit and five-qubit protocols shown in Fig.\ref{fig:scheme}(b), we can see that in the range $0.707 < p_H < 0.870$, the four-qubit protocol is more efficient, while beyond this range the five-qubit protocol has higher efficiency.
Hence once
$p_\text{H}(\rho_\text{int})$ reaches the optimal turning point
$p^\text{op}_\text{H}(\rho_\text{int})= 0.870$, the intermediate
state are then projected to $T$-direction by the twirling operation $D_\text{T}$ \cite{kitaevi05}.
$D_\text{T}$ converts the polarization from $H$-direction to the $T$-direction while deducing the polarization by a factor of $\sqrt{2/3}$.
Next these states are sent into the five-qubit $T$-type MSD module for the further
distillation. The hybrid protocol ultimately outputs almost pure
$T$-type magic states. The first criterion ($p_\text{T}(\rho_\text{in})>0.655$) is based on
the numerical result that in the region $\mathcal{A}_\text{T} \cap \mathcal{A}_\text{H}$, the five-qubit protocol is less efficient for only 1\% of the distillable states.
We can see that both ranges $\mathcal{A}_\text{T}$
and $\mathcal{A}_\text{H}$ are distillable by the hybrid protocol.
One interesting conclusion is that not just for $H$-type magic state,
the distillable range for the $T$-type magic state is also tight
in directions crossing the octahedron edges.

Compared with the seven-qubit MSD protocol,
the hybrid protocol can greatly reduce qubit cost for almost all of the distillable region.
Figure.\ref{fig:scheme}(b) shows that the seven-qubit protocol  performs with much lower efficiency in each round of distillation. Not only that,
the hybrid protocol has a considerable advantage in the necessary iteration number (Fig.\ref{fig:scheme}(c)).
For the region
$ \mathcal{A}_\text{H} - (\mathcal{A}_\text{T} \cap
\mathcal{A}_\text{H})$ (i.e.,
$p_\text{H}(\rho_\text{in}) > 0.707$ and $p_\text{T}(\rho_\text{in})
< 0.655)$, the qubit cost can
be evaluated as ${\left( {{4 \mathord{\left/
 {\vphantom {4 {{{\bar \theta }_4}}}} \right.
 \kern-\nulldelimiterspace} {{{\bar \theta }_4}}}} \right)^{{N_4}}} \cdot {\left( {{5 \mathord{\left/
 {\vphantom {5 {{{\bar \theta }_5}}}} \right.
 \kern-\nulldelimiterspace} {{{\bar \theta }_5}}}} \right)^{{N_5}}}$ for the hybrid protocol,
 while ${\left( {{7 \mathord{\left/
 {\vphantom {7 {{{\bar \theta }_7}}}} \right.
 \kern-\nulldelimiterspace} {{{\bar \theta }_7}}}} \right)^{{N_7}}}$ for the seven-qubit protocol. Here ${N}_{4(5)(7)}$ are the iteration numbers of the 4(5)(7)-qubit MSD scheme
and ${\bar \theta }_{4(5)(7)}$ are the average success probabilities (${\bar \theta }_4=0.244, {\bar \theta }_5=0.124$
and ${\bar \theta }_7=0.046$).
%Therefore, the consumption depends on two factors: the success
%probability  and  the necessary iteration number. Figure
%\ref{fig:scheme}(b) and (c) show the comparison between the hybrid
%MSD protocol and the previous MSD protocols. From Fig.
%\ref{fig:scheme}(b), we can obtain the average success probability
%for one iteration ${\bar \theta }_4=0.244, {\bar \theta }_5=0.124$
%and ${\bar \theta }_7=0.046$. From Fig. \ref{fig:scheme}(c), we can
%see that for the range  $\mathcal{A}_\text{H} -
%(\mathcal{A}_\text{T} \cap \mathcal{A}_\text{H})$ (i.e.,
%$p_\text{H}(\rho_\text{in}) > 0.707$ and $p_\text{T}(\rho_\text{in})
%< 0.655)$, the hybrid protocol has a considerable advantage in the
%necessary iteration number to achieve a $T$-type or $H$-type magic
%state with a target polarization above 0.999 (this corresponds to
%implementing one non-Clifford operation with theoretical fidelity
%0.9995 \cite{Cory02}).
The hybrid
protocol can reduce the qubit cost by a roughly
estimated factor of $10^{35}$ with respect to the seven-qubit MSD protocol \cite{SI},  with a target polarization above 0.999 (this corresponds to
implementing one non-Clifford operation with theoretical fidelity
0.9995 \cite{Cory02}).
The same observation can be extracted for the major part of the
region $\mathcal{A}_\text{T} \cap \mathcal{A}_\text{H}$ thanks to
the property of five-qubit protocol that the increase of the out polarization is exponentially fast in the number
of noisy copies when the polarization is high enough \cite{kitaevi05}.
Just for a little region around the pure magic state $\rho
_\text{H}$, the seven-qubit protocol can be slightly more efficient. The exceptional region occupies about 0.57\% of the distillable range.
%This proportion goes to zero when the target polarization goes to 1.

\begin{figure}
\centering \includegraphics[width=0.48\textwidth]{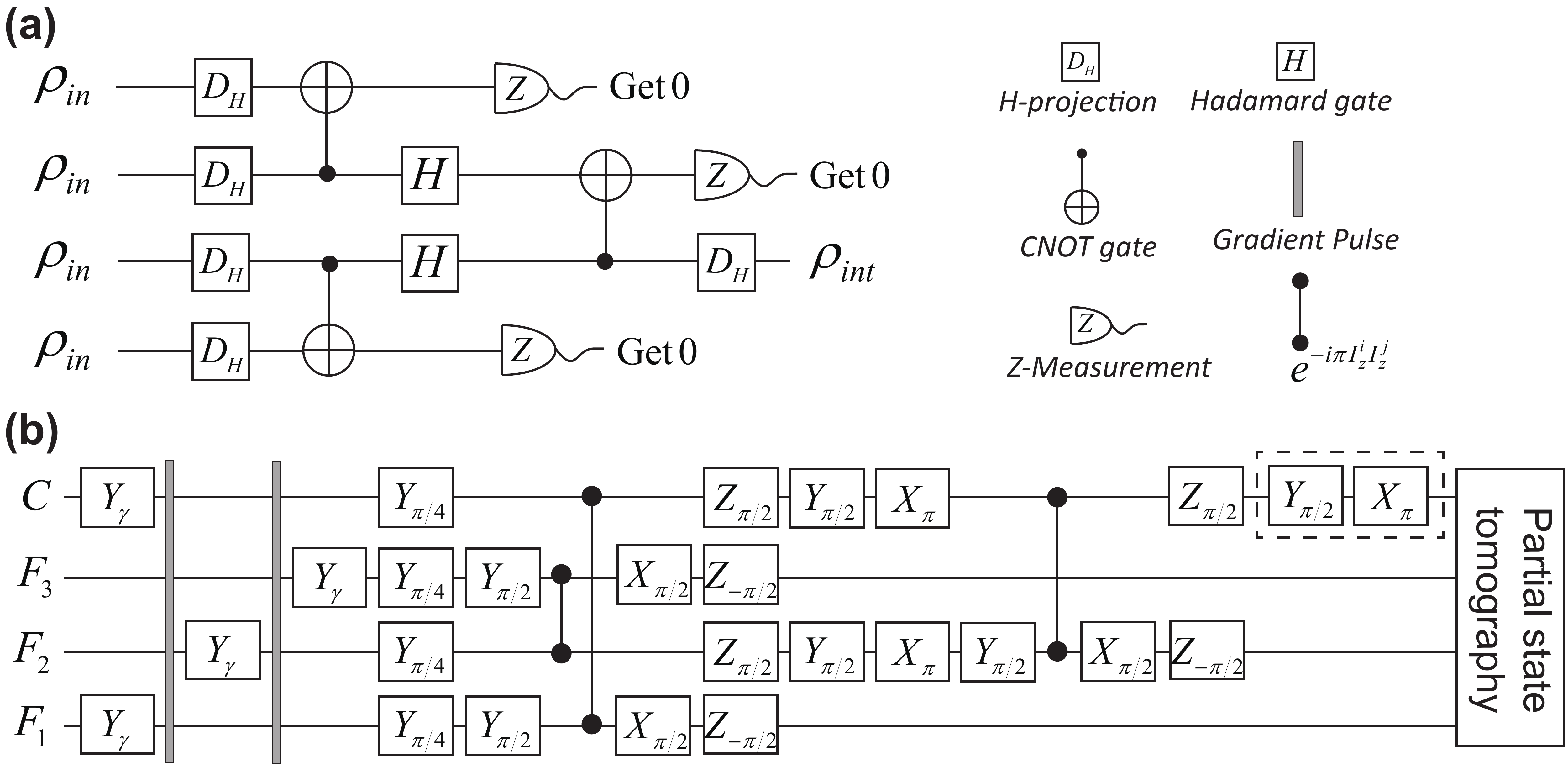} \caption{\label{fig:sequence}Quantum circuit (a) and the corresponding pulse sequence (b) for the four-qubit $H$-type MSD protocol,
where ${X_\alpha }{\rm{ = }}{e^{ - ia{I_x}}}$, ${Y_\alpha }{\rm{ = }}{e^{ - ia{I_y}}}$ and ${Z_\alpha }{\rm{ = }}{e^{ - ia{I_z}}}$ denote single-qubit rotations.
The $H$-projection operation ${D_\text{H}}\left( \rho  \right) = \left(\rho  + H\rho {H^\dag }\right)/2$ was realized by accumulating signals of two experiments: one with the Hadamard gate implemented by the pulses shown in the dashed box in (b), and the other without it.
}
\end{figure}

Now we look in detail at the four-qubit $H$-type MSD scheme, whose quantum circuit is shown in Fig. \ref{fig:sequence}(a): (i) first prepare four copies of a noisy magic state $\rho_\text{in}^{\otimes 4}$ as the input state; (ii) perform the parity-checking in pairs three times; (iii) if all measurements give $0$ result, i.e., the three measured qubits are in the state $\left| {000} \right\rangle $, the protocol succeeds and one applies the $H$-projection operation $D_H$ to the qubit that hasn't been measured. The output state is ${\rho _\text{dis}} \otimes |000\rangle \langle 000|$ with the success probability $\theta  =(2 + 2p_0^2 + p_0^4)/16$, where $\rho_\text{dis}$ has the output polarization:
\begin{equation}
{p_\text{H}}({\rho _\text{dis}}) = \frac{6 p_0^2 + p_0^4}{\sqrt 2 \left\{ 2 + 2 p_0^2 + p_0^4 \right\} }.
\end{equation}
It gives ${p_\text{H}}({\rho _\text{{dis}}}) > p_0$ when $0.7071 < p_0 < 0.9617$. Here $p_0 = {p_\text{H}}({\rho _\text{in}})$ is the initial polarization of the input states.

We experimentally demonstrate the four-qubit distillation protocol. The physical
system is iodotrifluroethylene ($\text{C}_2\text{F}_3\text{I}$) dissolved in $d$-chloroform. One ${}^{13}\text{C}$ nucleus and three ${}^{19}\text{F}$ nuclei are used as a four-qubit quantum information processor \cite{Peng2008,PengLuo2014}.
The natural Hamiltonian of the coupled spin system is ${\cal H}{\rm{ = }}\sum\nolimits_i {\cal H} _z^i + \sum\nolimits_{i < j} {\cal H} _c^{ij}$,
where ${\cal H}_z^i = \pi {\upsilon _i}{\sigma_z^i}$ is the Zeeman term, ${\upsilon _i}$ is the Larmor frequency of spin $i$, and ${\cal H}_{c}^{ij} = \left( {{\pi  \mathord{\left/ {\vphantom {\pi  2}} \right. \kern-\nulldelimiterspace} 2}} \right){J_{ij}}{\sigma_z^i}{\sigma_z^j}$ describes the interaction between spin $i$ and $j$, ${J_{ij}}$ is the scalar coupling strength. Experiments were performed on a Bruker AV- 500 spectrometer at room temperature. All of the relevant parameters along with the molecular structure are shown in the supplemental material \cite{SI}.

Figure \ref{fig:sequence}(b) shows the pulse sequence of the experiment, corresponding to the quantum circuit in Fig. \ref{fig:sequence}(a).  We first initialized the system in a pseudopure state (PPS) \cite{Chuang1997} ${\rho _{0000}} = (1 - \epsilon )\mathbf{1}/16 + \epsilon \left| {0000} \right\rangle \langle 0000|$ by using line-selective method \cite{Peng2001,PengLuo2014,SI}, where $\epsilon \approx10^{-5}$ is the
polarization.  Instead of first four $H$-projection operations in Fig. \ref{fig:sequence}(a), four copies of noisy $H$-type magic states were prepared by the depolarization procedure shown in the second box of Fig. \ref{fig:sequence}(b).
A rotation by an angle $\gamma$ around the $y$-axis transforms ${\sigma _z}$ to ${\sigma _z}\cos \gamma  + {\sigma _x}\sin \gamma $. The following gradient field destroys the $x$ component. By changing the rotation angle $\gamma$, we experimentally prepared five sets of noisy magic states, and each set has different average polarization:  ${p_\text{H}}( {\rho _\text{in}^1}) = 0.661, {p_\text{H}}( {\rho _\text{in}^2}) = 0.826, {p_\text{H}}( {\rho _\text{in}^3}) = 0.857, {p_\text{H}}( {\rho _\text{in}^4}) = 0.885, {p_\text{H}}( {\rho _\text{in}^5}) = 0.999.$

The three parity check gates of Fig. \ref{fig:sequence}(a) were implemented through the distillation procedure in Fig. \ref{fig:sequence}(b).
It consists of Clifford operations.
At the output side, the ${}^{13}\text{C}$ nucleus carries the distilled magic state $\rho_\text{dis}$.
To avoid the error accumulation and exhibit a near-perfect distillation step, we used one high-fidelity  shaped pulse gained by the gradient ascent
pulse engineering (GRAPE) algorithm \cite{Khaneja2005,Lu2011,Xu2012} to implement this sequence. The GRAPE pulses have durations of 16.8$ms$ with  theoretical fidelities above 0.996.

\begin{figure}[t]
\centering \includegraphics[width=0.48\textwidth]{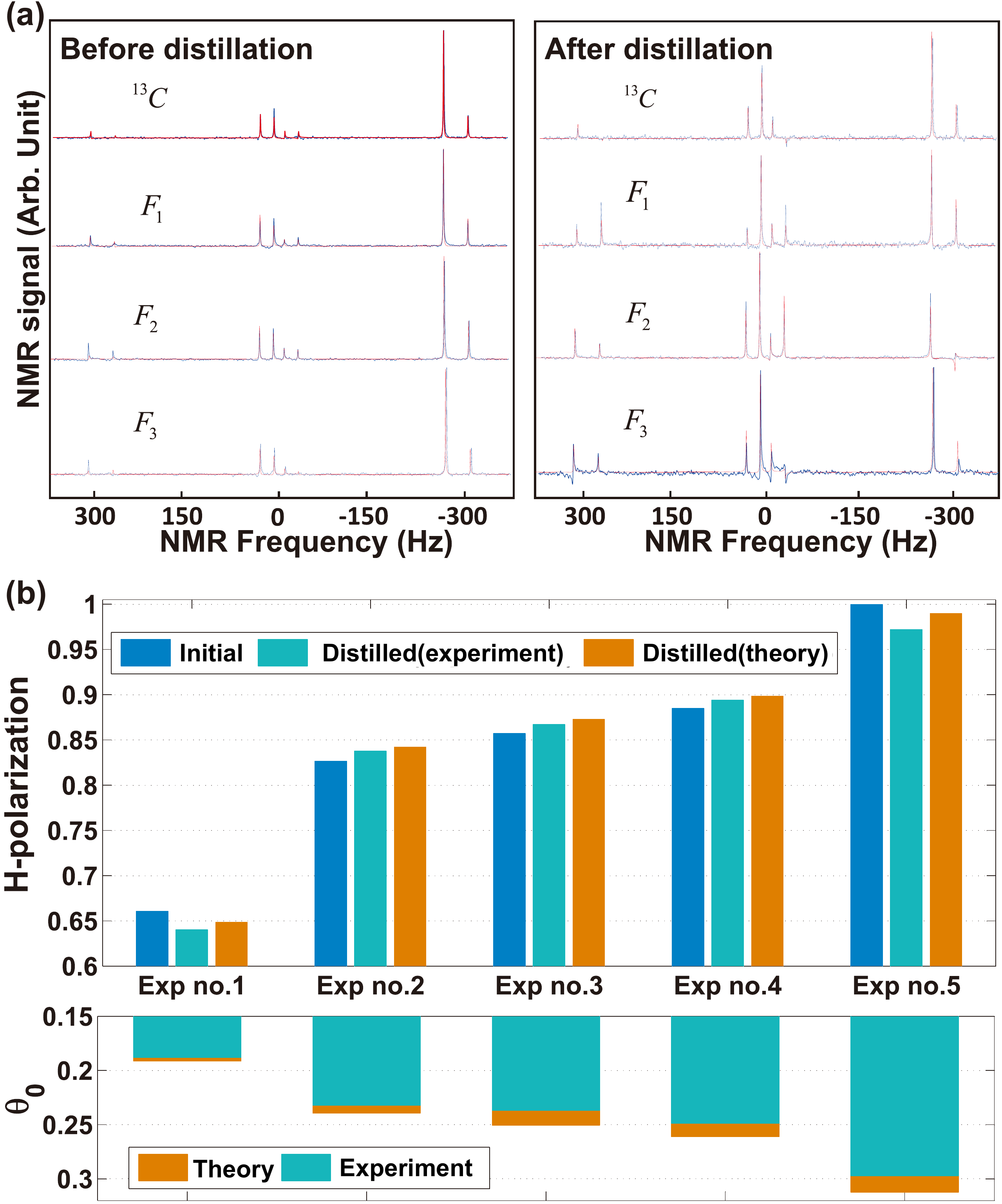} \caption{\label{fig:result}
(a) Experimental results of the distillation. Upper: polarizations along $H$-direction before and after distillation. Lower: the success probability $\theta_0$ of distillation versus to the input polarization. (b) Experimental spectra of the four nuclei before and after distillation for $p_\text{H}(\rho_\text{in}) = 0.826$. The experimentally measured and simulated spectra are shown as the blue and red curves, respectively.
}
\end{figure}

The distilled output state $\rho_\text{dis}$ can be written as (the ${}^{13}\text{C}$ nucleus is labeled as qubit 1)
\begin{equation}
\begin{split}
&{\rho _\text{out}^\text{expt}} = \sum\limits_{i = 0}^7 {{\theta _i}} {\rho _i} \otimes \left| i \right\rangle \left\langle i \right| + \sum\limits_{i \ne j = 0}^7 {{\theta _{ij}}} {\rho _{ij}} \otimes \left| i \right\rangle \left\langle j \right|,\\
&{\rho _i} = \frac{1}{2}\left( {\mathbf{1} + {x_{i}}{\sigma _x} + {z_{i}}{\sigma _z}} \right),
\end{split}
\end{equation}
where ${{\theta _i}}$ is the probability of the measurement outcome,
corresponding to the resulting state $\left| i \right\rangle $ of
the other three redundant qubits, and $\left| i \right\rangle  =
\left| {000} \right\rangle ,\left| {001} \right\rangle ,...,\left|
{111} \right\rangle $, for $i = 0,1,2,...,7$. Measuring the outcome
$\left| {000} \right\rangle $ indicates a successful purification.
In NMR quantum information processing, since only ensemble
measurements are available, we directly measure the expectation
value of an observable, without projective measurements. In
spite of this, we can obtain all $\theta_i$, $x_i$ and $z_i$ using
partial quantum state tomography \cite{Lee2002} to see the
purification effect. Five readout operations are sufficient to
determine all the wanted parameters. Firstly, by directly reading the
signal of ${}^{13}\text{C}$, we can obtain all ${\theta _i}{x_{i}}$;
secondly, by reading the signal of ${}^{13}\text{C}$  after the
application of a ${\pi  \mathord{\left/{\vphantom {\pi  2}}
\right.\kern-\nulldelimiterspace} 2}$ pulse, we can get all ${\theta
_i}{z_{i}}$; the additional three readout operations consist of
applying a ${\pi  \mathord{\left/{\vphantom {\pi  2}}
\right.\kern-\nulldelimiterspace} 2}$ pulse on $\text{F}_1$,
$\text{F}_2$ and $\text{F}_3$, then reading the signals. The spectra
of the four nuclei after applying a ${\pi  \mathord{\left/{\vphantom
{\pi  2}} \right.\kern-\nulldelimiterspace} 2}$ pulse are shown in
Fig. \ref{fig:result}(a). They  are sufficient to determine all
diagonal terms  of the density matrix, that means we obtain all
${\theta _i} = {m_{i+1}} + {m_{i + 9}}$, where $m_i$ represents the
$i$th diagonal term. Since this sample is unlabeled, we must
transfer the polarizations of the ${}^{19}\text{F}$ spins to the
${}^{13}\text{C}$ spin by SWAP gates and then read the information
of the ${}^{19}\text{F}$ spin through the ${}^{13}\text{C}$ spectrum
\cite{Peng2010}. The experimental results are shown in Fig.
\ref{fig:result}(b). The corresponding measured output polarizations
are ${p_\text{H}}( {\rho _\text{out}^1}) = 0.640, {p_\text{H}}(
{\rho _\text{out}^2}) = 0.838, {p_\text{H}}( {\rho _\text{out}^3}) =
0.867, {p_\text{H}}( {\rho _\text{out}^4}) = 0.894, {p_\text{H}}(
{\rho _\text{out}^5}) = 0.979$.  We see that the H-polarization of the noisy magic states
$\rho _\text{in}^2,\rho _\text{in}^3,\rho _\text{in}^4$ have been
experimentally improved by the four-qubit MSD protocol because their
input polarizations are in the distillable range. It shows that one can enchance the $H$-polarization of the quantum states in $\mathcal{A}_\text{H}$ with initial $T$-polarization $p_\text{T}\leq0.655$ to $p_\text{H}\geq0.870$ by the four-qubit MSD, and send them into the next step of the hybrid protocol, five-qubit MSD, for converging to the pure magic state $\rho _\text{T}$.

The total experimental time of the distillation procedure including the readout procedure ranges from $25ms$ to $40ms$. It is short compared to the transversal relaxation times
of the nuclei (the minimum $T_2^*$ of the four nuclei is about $250ms$), so the signal attenuation caused by the spin-spin relaxation effect is small. We numerically optimized all GRAPE pulses so that they are robust to 5\% inhomogeneity of the r.f. field.
By doing this, the influence of the r.f. field inhomogeneity is largely eliminated. For the four input copies, the biggest polarization deviation of the individual spin from the average polarization is 0.029. A detailed numerical analysis on the robustness of the distillation algorithm to these differences of input polarizations is presented in the supplemental material \cite{SI}. It shows that the distillation algorithm is strongly robust to the imperfect copies of the initial state. The relative deviations between the experimental results and the theoretical expectation are $0.6\% \sim 1.2\%$. They mainly come from the imperfections of GRAPE pulses, experimental parameters and data processing.

In conclusion, we presented a hybrid MSD protocol, which aims at
taking advantage of different MSD protocols.
It further integrates all of the currently known distillable ranges and
extends the $T$-type distillable range to the stabilizer octahedron
edges. Moreover, the hybrid scheme is optimized in efficiency
and has a remarkable advantage in saving qubit resources. The hybrid
construction exhibits the ability to establish a unified framework
for different MSD protocols. It shows that if a state provides quantum computation power in either $H$- or $T$- direction, it can keep the power in the other magic direction by stabilizer operations. We
also experimentally demonstrate the four-qubit scheme by the NMR
technology. The present experimental results, together with the
previous NMR experiment for the five-qubit protocol, confirm the
feasibility of the hybrid MSD scheme. It is expected that as more
MSD methods are put forward, more and more distinguished
combinations will come out according to the hybrid formalism.

\section{Acknowledgments}

Y. Yu thanks to get in the topic of ``magic state distillation" when she stayed in Laflamme's group.
This work is supported by the National
Key Basic Research Program of China (Grant No.
2013CB921800), National Natural Science Foundation of
China under Grant Nos. 11375167, 11227901, 91021005,
the Chinese Academy Of Sciences, the Strategic Priority
Research Program (B) of the CAS (Grant No.
XDB01030400), Research Fund for the Doctoral Program
of Higher Education of China under Grant No.
20113402110044 and the Scientific Research Foundation
for the Returned Overseas Chinese Scholars, State Education
Ministry.

%\bibliographystyle{apsrev4-1}
%\bibliography{ref}

\section{Supplementary Materials}

\subsection{1. The performance of the four-qubit protocol in the asymptotic regime}

The iterative function of the four-qubit protocol is
\begin{equation}
{\varepsilon _{out}} = f\left( \varepsilon  \right) = 0.5{\rm{ - }}\frac{{6{{\left( {1{\rm{ - }}2\varepsilon } \right)}^2}{\rm{ + }}{{\left( {1{\rm{ - }}2\varepsilon } \right)}^4}}}{{\sqrt 8 \left[ {2 + 2{{\left( {1{\rm{ - }}2\varepsilon } \right)}^2}{\rm{ + }}{{\left( {1{\rm{ - }}2\varepsilon } \right)}^4}} \right]}},
\end{equation}
where $\varepsilon  = \frac{{1 - {p_H}}}{2} $ is the error probability. Its first-order Taylor expansion near the polarization ${p_H} = 0.962$ (i.e. $\varepsilon  = 0.019$) is
\begin{equation}
\begin{split}
{\varepsilon _{out}} &= f\left( {0.019} \right) + f'\left( {0.019} \right)\left( {\varepsilon  - 0.019} \right) + o\left[ {{{\left( {\varepsilon  - 0.019} \right)}^2}} \right] \\
                     &= 0.019 + 0.75\left( {\varepsilon  - 0.019} \right) + o\left[ {{{\left( {\varepsilon  - 0.019} \right)}^2}} \right].
\end{split}
\end{equation}
${\varepsilon ^*} = 0.019$ is the convergence value, and the convergence rate is linear, which is different from the five-qubit protocol's quadratic convergence in the asymptotic regime.
For the input ${\varepsilon _0}$, after the first iteration, ${\varepsilon _1} \approx 0.019 + 0.75\left( {{\varepsilon _0} - 0.019} \right)$. After the second iteration, ${\varepsilon _2} \approx 0.019 + {0.75^2}\left( {{\varepsilon _0} - 0.019} \right)$. After $k$ times of iterations, ${\varepsilon _n} \approx 0.019 + {0.75^k}\left( {{\varepsilon _0} - 0.019} \right)$.
Near the ${p_H} = 0.962$, we can approximate the success probability ${\theta _s} \approx \theta \left( {0.962} \right) = 0.294$.
The total number of initial noisy magic states $N = {\left( {{4 \mathord{\left/
 {\vphantom {4 {{\theta _s}}}} \right.
 \kern-\nulldelimiterspace} {{\theta _s}}}} \right)^n} \approx {\left( {13.6} \right)^k}$. So
 \begin{equation}
\begin{split}
 {\varepsilon _{out}} &\approx 0.019 + {0.75^{{{\log }_{13.6}}N}}\left( {{\varepsilon _0} - 0.019} \right) \\
                      &  = 0.019 + {N^{\left( {\frac{1}{{{{\log }_{0.75}}13.6}}} \right)}}\left( {{\varepsilon _0} - 0.019} \right).
 \end{split}
\end{equation}
The error rate in the distilled magic states is reduced polynomially with respect to the number of noisy input magic states.

Nevertheless, in our hybrid scheme,
before the polarization reaches the asymptotic regime of the four-qubit protocol, we switch to the  five-qubit protocol which gives an exponential decay of the error rate.
The four-qubit protocol shows the ability of reducing the qubit cost mainly rooted in its less qubit cost in every round of distillation, which brings about the higher efficiency than five-qubit protocol before the optimal turning point ($p_H=0.870$).
%The efficiency is defined as $\Delta {p_{T\left( H \right)}}\theta /n$, which is the parameter in comparing various protocols in our work.

\subsection{2. The integration between the 4-qubit $H$-type MSD and the 5-qubit $T$-type MSD }

In the hybrid magic state distillation (MSD) scheme,
the noisy copies firstly enter into $H$-type MSD modules as four qubits in one group. With the increase of iterations number,
the modules output states more closer to $H$-type magic state. Certainly, their polarizations of $T$-direction $p_T$ also
increase. Once $p_H$ is higher than the optimal turning point ($p_H=0.87$,corresponding to $p_T=0.71$), the state enters into $T$-type
MSD module. Then after several iterations of $T$-type MSD, we obtain states with $p_T \sim 1$. Fig. \ref{fig:scheme} shows the integration
between the two schemes.

\begin{figure*}[ht]
\centering \includegraphics[width=0.99\textwidth]{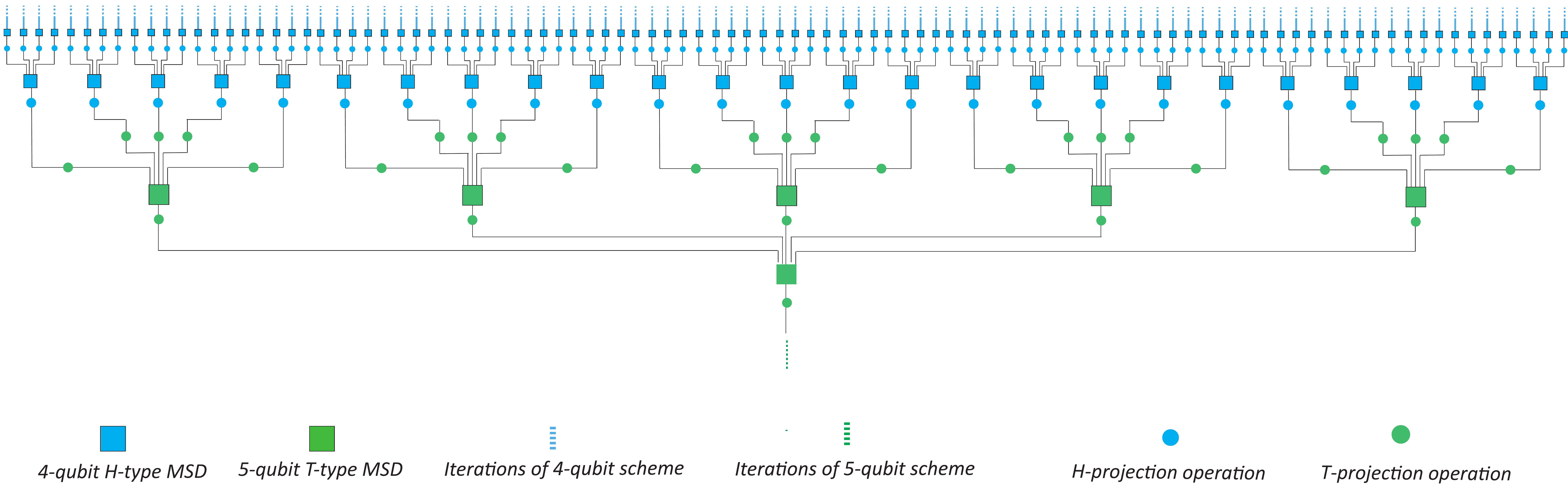} \caption{\label{fig:scheme}
The integration between the 4-qubit $H$-type MSD and the 5-qubit $T$-type MSD.
}
\end{figure*}

\subsection{3. The qubit cost in the low polarization range $ \mathcal{A}_\text{H} - (\mathcal{A}_\text{T} \cap
\mathcal{A}_\text{H})$}

\begin{figure}[h]
\centering \includegraphics[width=0.48\textwidth]{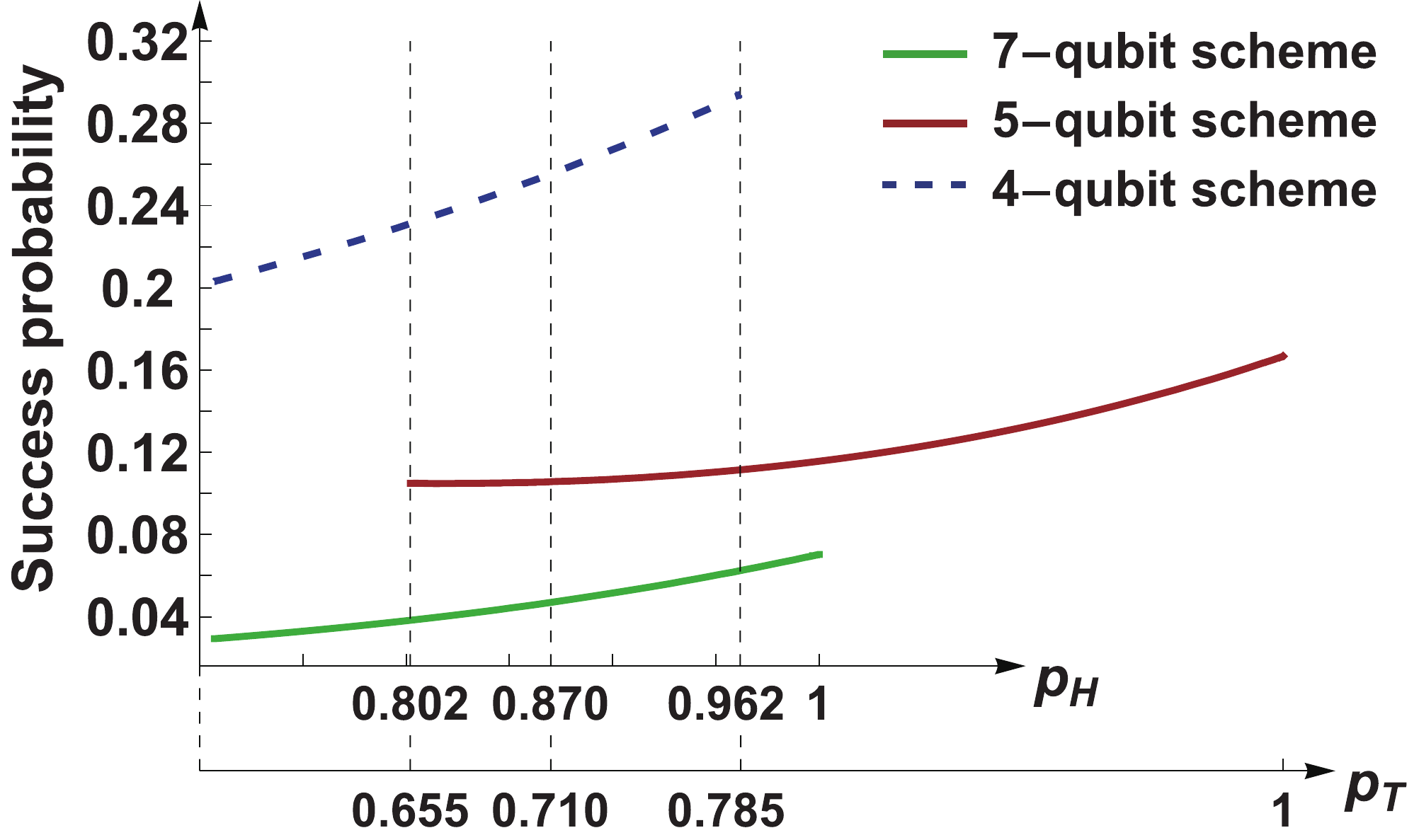} \caption{\label{fig:probability}
The success probability of the three individual MSD protocols.
}
\end{figure}

From Fig. \ref{fig:probability}, we can obtain the average success probability for one iteration (${\bar \theta }_4 = 0.244$, ${\bar \theta }_5 = 0.124$
and ${\bar \theta }_7 = 0.046$).
For the region $ \mathcal{A}_\text{H} - (\mathcal{A}_\text{T} \cap
\mathcal{A}_\text{H})$, the necessary iteration number to achieve a $T$-type or $H$-type magic
state with a target polarization above 0.999 is about 26 using the seven-qubit protocol. The  qubit cost is evaluated as ${\left( {{7 \mathord{\left/
 {\vphantom {7 {{{\bar \theta }_7}}}} \right.
 \kern-\nulldelimiterspace} {{{\bar \theta }_7}}}} \right)^{{26}}} \sim 10^{56}$. Using the hybrid protocol, it needs about 11 iterations of the four-qubit protocol and 5 iterations of the
 five-qubit protocol. The  qubit cost is evaluated as ${\left( {{4 \mathord{\left/
 {\vphantom {4 {{{\bar \theta }_4}}}} \right.
 \kern-\nulldelimiterspace} {{{\bar \theta }_4}}}} \right)^{{11}}} \cdot {\left( {{5 \mathord{\left/
 {\vphantom {5 {{{\bar \theta }_5}}}} \right.
 \kern-\nulldelimiterspace} {{{\bar \theta }_5}}}} \right)^{{5}}} \sim 10^{21}$. Hence the qubit cost is reduced by a factor of about $10^{35}$ using the hybrid protocol for the low polarization range $ \mathcal{A}_\text{H} - (\mathcal{A}_\text{T} \cap
\mathcal{A}_\text{H})$.

As an example, table \ref{fig:compare} shows the performance of the hybrid protocol and the 7-qubit protocol, when they are used to distill one noisy state whose polarization is in $H$-direction ($p_H=0.78$,
equivalent to $p_T = 0.636$). Though the target states of the two schemes are different ($p_H = 0.99$ in the 7-qubit case,
$p_T = 0.999$ in the hybrid case), either of the target states can be used to implement one non-Clifford operation with
theoretical fidelity 0.9995. We can see that the hybrid protocol requires less iterations and possesses much higher successful probability than the 7-qubit protocol, which lead to the great saving in
qubit cost.

\begin{table}[h]
\centering \includegraphics[width=0.48\textwidth]{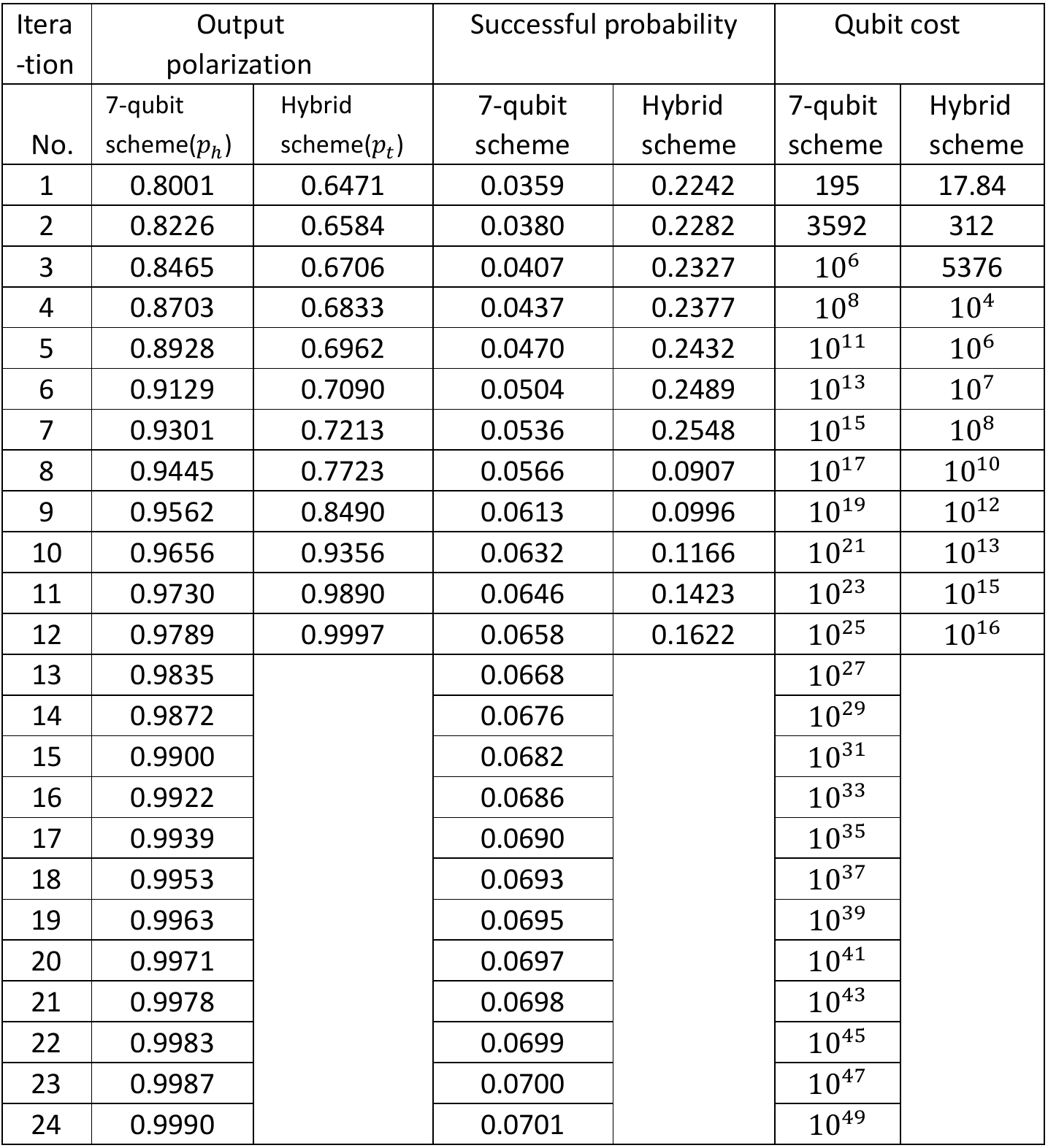} \caption{\label{fig:compare}
Comparing the performance of the hybrid MSD protocol and the 7-qubit protocol,  when they are used to distill one noisy state of $H$-direction with $p_H=0.78$.
Since the target direction of the hybrid protocol is $T$-direction, the polarizations gained by the subprocedure of the 4-qubit $H$-type protocol have been converted to $T$-direction, i.e., multiplied by the factor $\sqrt{2/3}$.
}
\end{table}

\subsection{4. The efficiency of the whole procedure}
\begin{figure}[h]
\centering \includegraphics[width=0.48\textwidth]{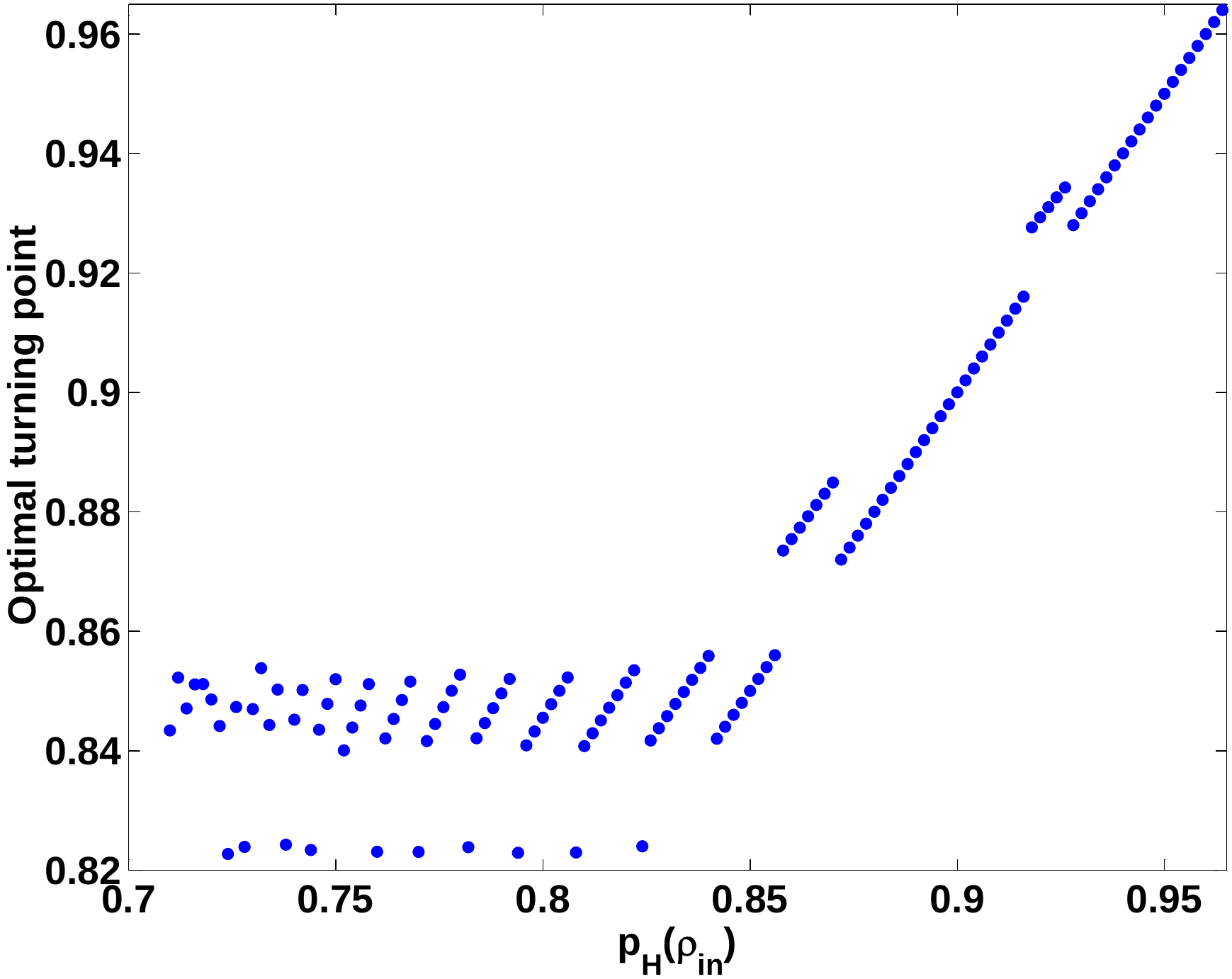} \caption{\label{fig:op}
The optimal turning points for different input polarizations with the
target polarization above 0.999. The optimal turning points gather around $p_H=0.85$. If the input state's polarization is higher than 0.86, it's better to distill them directly using the five-qubit $T$-type MSD protocol.
}
\end{figure}

The parameter $\nu = \Delta {p_{T\left( H \right)}}\theta /n$ represents the increased polarization per consumed qubit in each iteration. The hybrid protocol is optimized by $\nu$. The efficiency of the whole procedure of $k$ times distillation is defined as
 \begin{equation}
V = \frac{{\Delta {p_1} + \Delta {p_2} +  \cdots  + \Delta {p_k}}}{{{{{n_1}{n_2} \cdots {n_k}} \mathord{\left/
 {\vphantom {{{n_1}{n_2} \cdots {n_k}} {{\theta _1}{\theta _2} \cdots {\theta _k}}}} \right.
 \kern-\nulldelimiterspace} {{\theta _1}{\theta _2} \cdots {\theta _k}}}}}.
 \end{equation}
Both $\nu$  and $V$ is determined by two parameters: increment $\Delta {p_{T\left( H \right)}}$ gained from one iteration and the average qubit consumption $n/\theta$ in one iteration.
Given a target polarization, the optimal turning point may be slightly different from the one calculated by the efficiency $\nu$. Setting the target $P_T=0.999$,
for different initial polarizations,  we numerically calculated the optimal turning points, which are shown in Fig. \ref{fig:op}. We can see the optimal turning points gather around $p_H=0.85$,
which is slightly different from the hybrid protocol's turning points $p_H=0.87$.

\subsection{5. The hybrid protocol}

Once the polarization reaches the turning point $p_H=0.87$ of the hybrid protocol, the $T$-projection operation $D_T$ is performed to the state. Then we get a polarization of $T$-direction with $P_T>0.71$.

1. If $p_T(\rho_{in})>0.71$, we directly distill them with five-qubit protocol.

2. If $0.655<p_T(\rho_{in})<0.71$ and $p_H(\rho_{in})<0.707$, the state can't benefit from the four-qubit protocol. We should distill it with the five-qubit protocol.

3. If $p_T(\rho_{in})<0.655$ and $0.707<p_H(\rho_{in})$, the state can't directly benefit from the five-qubit protocol. We should firstly distill it to $p_H(\rho_{in})=0.87$ with the four-qubit protocol.

4. If $0.655<p_T(\rho_{in})<0.71$ and $0.707<p_H(\rho_{in})<0.87$, for about $95\%$ of the states in this range, it's better to directly distill them with the five-qubit protocol.

\subsection{6. Robustness of the 4-qubit $H$-type MSD algorithm}

In the theoretical analysis, the input of the 4-qubit $H$-type MSD algorithm is four qubits with the same polarization of $H$-direction.  However, it is impossible to prepare each qubit to totally identical state. It is unavoidable that there are some differences between the polarizations of the four copies in the experiment.
Table. \ref{tab:initial} shows the initial polarizations in our experiment.
We can see that the biggest polarization deviation of the
individual spin from the average polarization is 0.029.
It is important to analyze the robustness of the distillation algorithm to these differences of polarizations of
input states.

\begin{table}[h]
\includegraphics[scale=0.5]{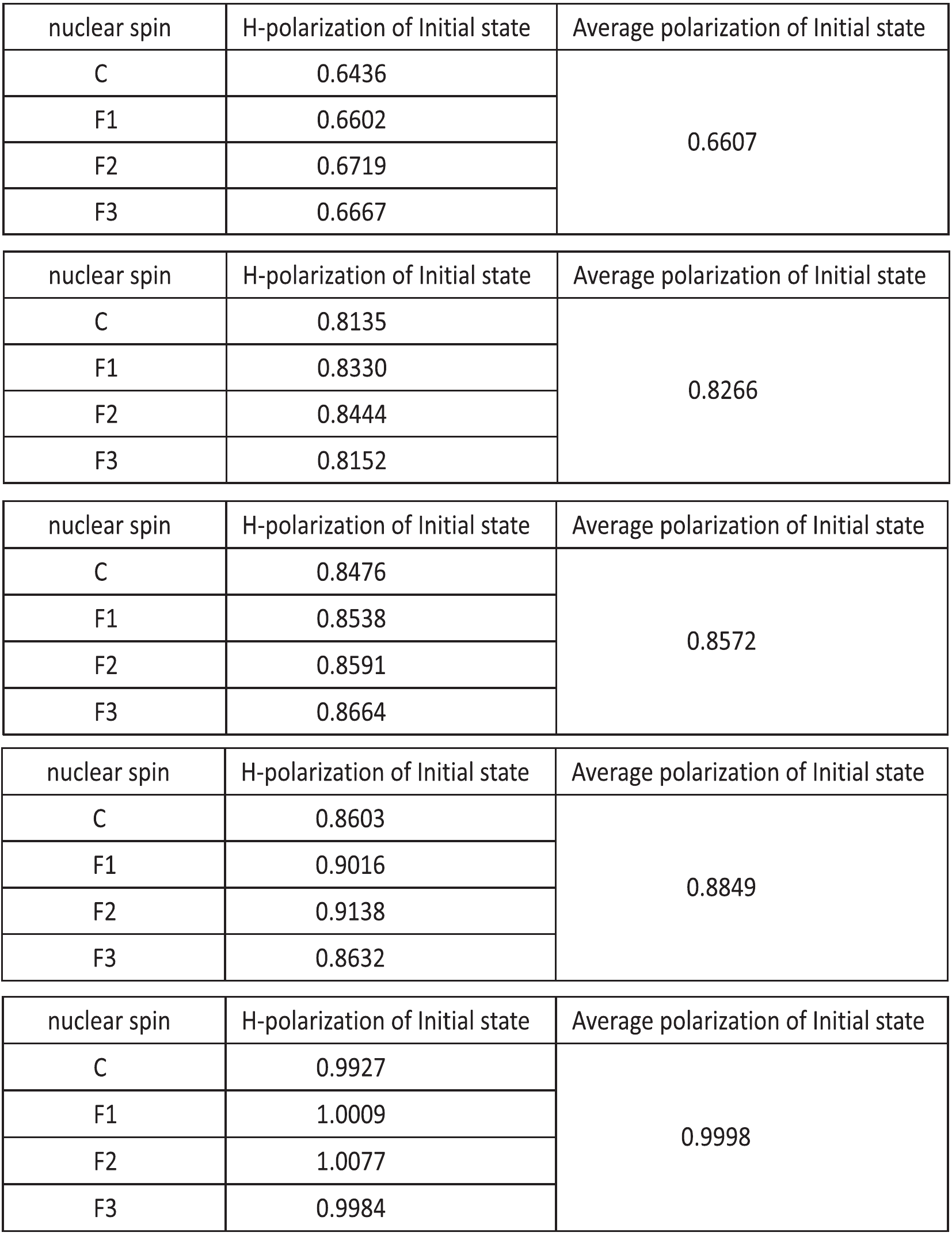}
\caption{\label{tab:initial}  The polarizations of the prepared initial states. }
\end{table}

\begin{figure*}[ht]
\centering \includegraphics[width=0.95\textwidth]{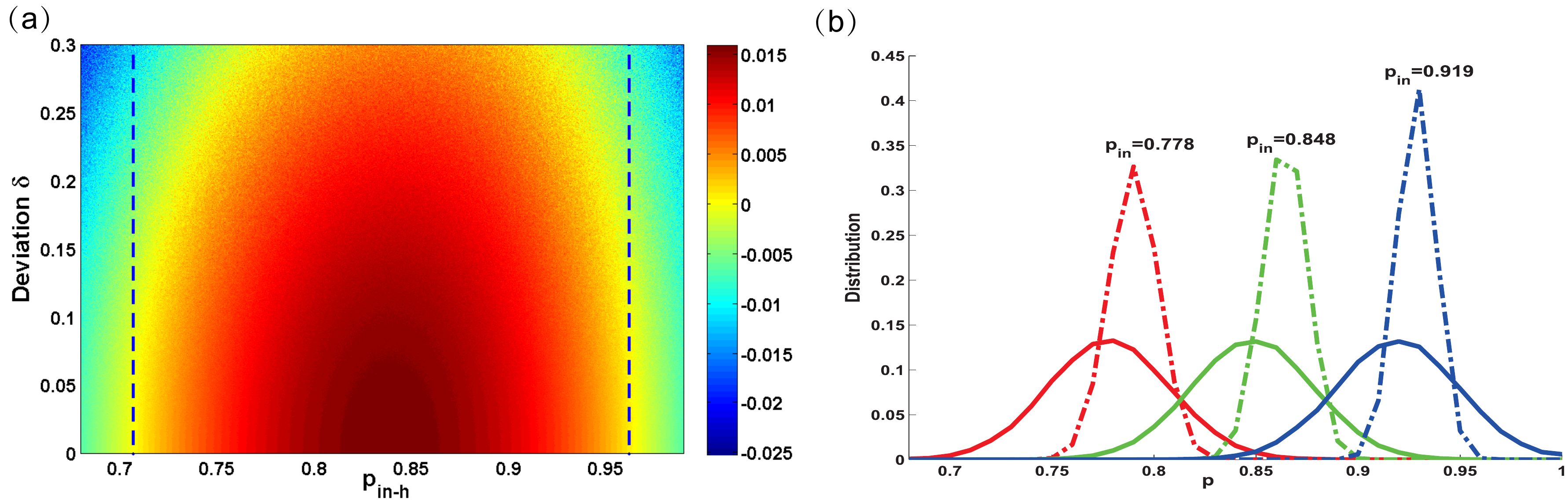} \caption{\label{fig:robust}
(a)Distillation effect versus to average input polarization and the deviations of input polarizations from the average value.
For example, the point (0.82, 0.08) in the $\delta$  $\&$ ${p_{in}}$ plane, means that we randomly choose the input polarizations from
the range 0.82-0.08 to 0.82+0.08. The blue perpendicular lines  represent the theoretical  distillable boundary.
(b)Distributions of output polarization correspond to gaussian distribution inputs.
The solid curves represent the input polarization of gaussian distributions with centre at 0.778, 0.848 and 0.919. The dashed curves show the distributions of output.
}
\end{figure*}

Figure. \ref{fig:robust}a shows the average distillation effect versus to both the input polarization and the deviation. For every centre point which range from 0.68 to 0.99, we choose the deviations from 0 to 0.3. For each deviation, we randomly choose 100 inputs, then calculate the value $p_{out}-p_{in}$ for each input. We use the average of $p_{out}-p_{in}$ to evaluate the purity effect of the algorithm at this centre point. It shows that this distillation algorithm is strongly robust to the differences of polarization between the input qubits.
Figure. \ref{fig:robust}b shows the output polarization corresponding to gaussian distribution input. We can see the distributions of output present larger average and smaller variance comparing to input distribution.

%\begin{figure}
%\centering \includegraphics[width=0.48\textwidth]{guass.pdf} \caption{\label{fig:guass}
%Distributions of output polarization correspond to gaussian distribution inputs.
%The solid curves represent the input polarization of gaussian distributions with centre at 0.778, 0.848 and 0.919. The dashed curves show the distributions of output.
%}
%\end{figure}

%\bigskip

\subsection{7. Sample information}

\begin{figure}[h]
\centering \includegraphics[width=0.5\textwidth]{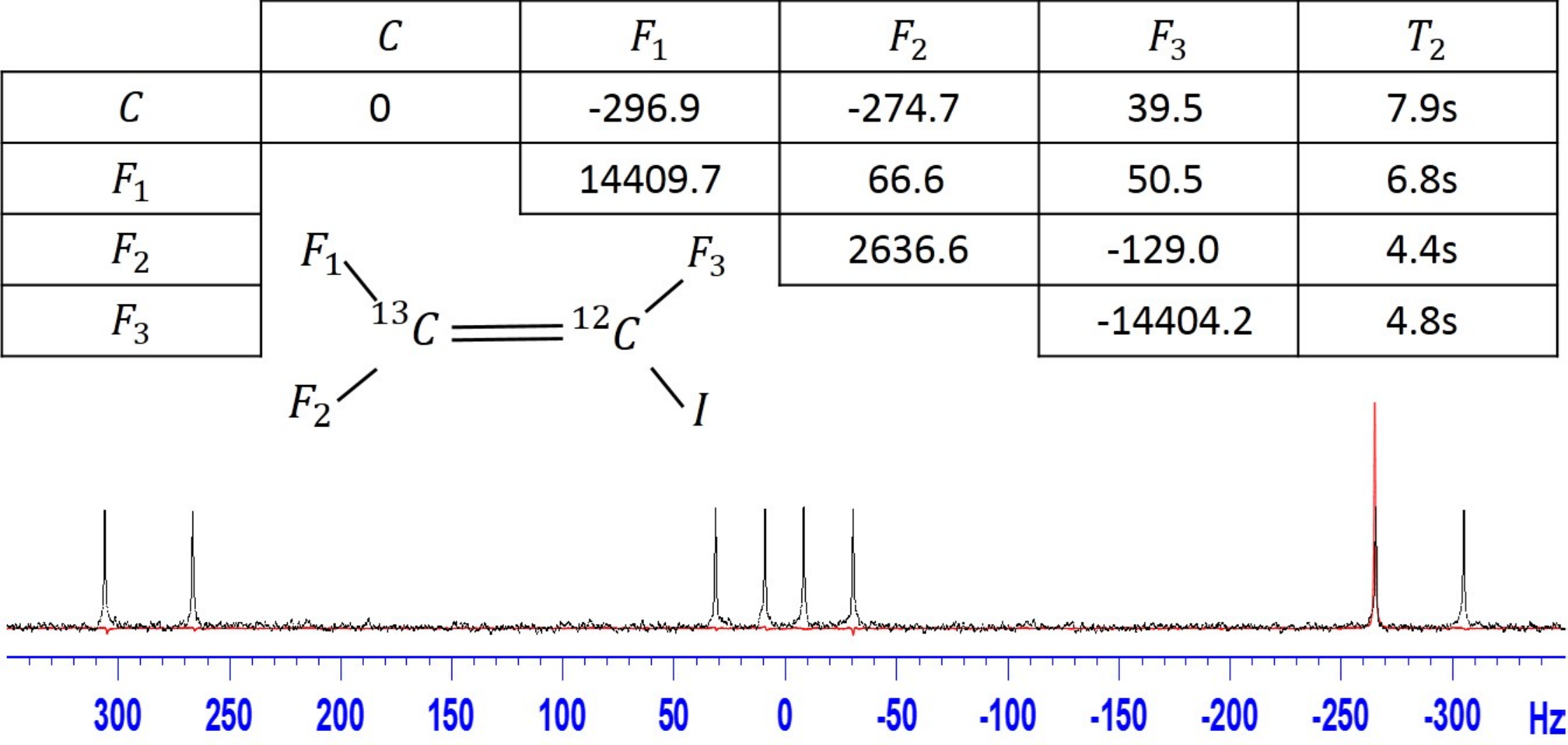} \caption{\label{fig:structure} Characteristics of the four-qubit quantum register.
The inset shows the structure, where the four qubits are labeled as ${}^{13}C$, $F_1$, $F_2$, $F_3$.
The chemical shifts and scalar coupling constants (in Hz) are on and above the diagonal in the table, respectively. The last column shows the transversal relaxation time
${T_2}$ of each nucleus measured by CPMG sequences. Shown below are spectra of ${}^{13}C$ obtained by $\pi /2$ readout pulses when the system is prepared in the thermal equilibrium state (black) and PPS ${\rho _{0000}}$ (red).
}
\end{figure}

The physical
system we used is the molecules of Iodotrifluroethylene ($C_2F_3I$) dissolved in d-chloroform. As ${}^{13}C$ and ${}^{19}F$ are spin-$\frac{1}{2}$ nuclei,
four qubits can be encoded using this sample for NMR quantum information processing.  The natural Hamiltonian of the coupled spin system is, ${\cal H}{\rm{ = }}\sum\nolimits_i {\cal H} _i^z + \sum\nolimits_{i < j} {\cal H} _{ij}^c$, where ${\cal H}_i^z = \pi {\upsilon _i}{\sigma_z^i}$ is the Zeeman Hamiltonian, ${\upsilon _i}$ is the Larmor frequency of spin $i$, and ${\cal H}_{ij}^c = \left( {{\pi  \mathord{\left/ {\vphantom {\pi  2}} \right. \kern-\nulldelimiterspace} 2}} \right){J_{ij}}{\sigma_z^i}{\sigma_z^j}$ describes the interaction between spin $i$ and $j$, ${J_{ij}}$ is the scalar coupling strength. All of the relevant parameters along with molecular structure are shown in Fig. \ref{fig:structure}.

%\bigskip

\subsection{8. PPS preparation}
The system is originally in the thermal equilibrium state ${\rho _{eq}} = {\rm I}/16 + \sum\limits_{i = 1}^4 {{\varepsilon _i}} I_z^i$, where  ${\varepsilon _i}\sim{10^{ - 5}}$ and $\bm{I}=(I_x, I_y, I_z)$ is the spin vector operator. For PPS ${\rho _{0000}} = (1 - {_0}){\rm I}/16 + {_0}\left| {0000} \right\rangle \langle 0000|$, the populations of all energy levels must be equalized except the first level. For this purpose, we numerically found an array ${\left\{ {{x_\alpha}} \right\}_{\alpha = 1,...14}}$, which determines a unitary operator
\begin{equation}
{U_1} = \sum\limits_\alpha {\exp \left[ { - i{x_\alpha}I_x^{\left( {\alpha + 1,\alpha + 2} \right)}} \right]},
\end{equation}
${I_x^{\left( {\alpha + 1,\alpha + 2} \right)}}$ is the single quantum transition operator between levels $\alpha+1$ and $\alpha+2$. The ${U_1}$ satisfies following requirement:
$diag[{U_1}{\rho _{eq}}U_1^\dag ] = diag[{\rho _{0000}}]$. That is, this unitary operator achieves saturation of latter 15 energy levels, while the population of the first level keeps unchanged. Then one gradient field pulse destroys all the coherences except homonuclear zero coherences of ${}^{19}F$ nuclei. The other specially designed unitary operator $U_2$ applies to the system, which transforms these redundant zero coherences to others that can be eliminated by gradient field pulse. Then applying another gradient field pulse, we prepare the PPS $\rho _{0000}$.
As $U_1$ is obtained by numerical search and $U_2$ is actually a combination of some CNOT gates between two selected levels, it is hard to find out a conventional pulse sequence to implement these two unitary operators. We engineered each operator as an individual shaped pulse by the gradient ascent pulse engineering (GRAPE) algorithm \cite{Khaneja2005}. These two GRAPE pulses are of the duration around 25ms with the theoretical fidelity above 0.994 and they are also designed with robustness against the rf inhomogeneity.

%\bigskip

\subsection{9. Experimental results of 4-qubit H-type MSD}

Table \ref{xout} shows the experimental results 4-qubit H-type MSD. The middle three inputs are in the distillable range. We observe higher polarizations corresponding to these three inputs.

\begin{table}[H]
\centering
\includegraphics[scale=0.46]{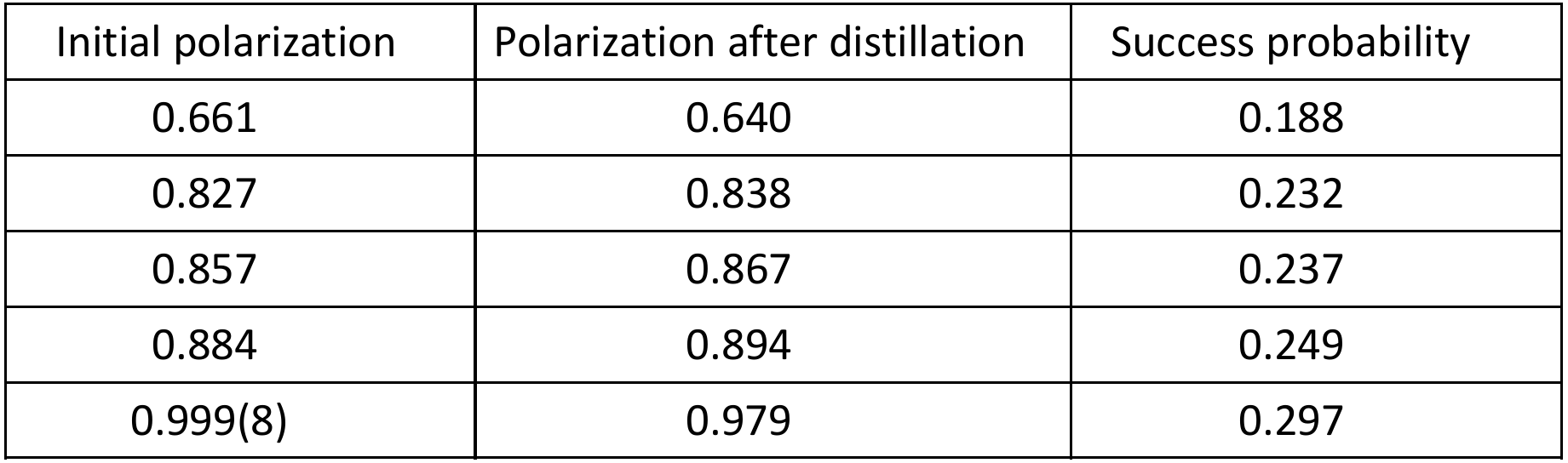}
\caption{\label{xout} Experimental results of 4-qubit H-type MSD. }
\end{table}

\subsection{10. State tomography}

\subsubsection{Input state tomography}

The input state can be written as
\begin{equation*}
\begin{split}
 {\rho _0} &= \frac{1}{2}\left[ {I + {x_{\rm{1}}}({\sigma _x} + {\sigma _z})} \right] \otimes \frac{1}{2}\left[ {I + {x_{\rm{2}}}({\sigma _x} + {\sigma _z})} \right] \otimes  \cdots  \\
 &\quad  \frac{1}{2}\left[ {I + {x_{\rm{3}}}({\sigma _x} + {\sigma _z})} \right] \otimes \frac{1}{2}\left[ {I + {x_{\rm{4}}}({\sigma _x} + {\sigma _z})} \right]\\
 &{\rm{ = }}\sum\limits_{i = 0}^7 {\sum\limits_{j = 0}^7 {\frac{1}{2}\left[ {I + {x_{\rm{1}}}({\sigma _x} + {\sigma _z})} \right]} }  \otimes {m_{ij}}\left| i \right\rangle \langle j|, \\
% &\sum\limits_{k = 0}^7 {{m_{kk}}}  = 1,\left| i \right\rangle ,\left| j \right\rangle  = \left| {000} \right\rangle ,\left| {001} \right\rangle ,...,\left| {111} \right\rangle.
 \end{split}
\end{equation*}
where $\sum\limits_{k = 0}^7 {{m_{kk}}}  = 1,\left| {i\left( j \right)} \right\rangle  = \left| {000} \right\rangle ,\left| {001} \right\rangle ,...,\left| {111} \right\rangle$. So we can get $x_1$ by summing all signals of qubit 1. Similarly, we can get all $x_i$. The average of $x_i$ is viewed as the polarization $x_{in}$ of input state.

\subsubsection{Distilled state tomography}

The state after distillation can be written as (assuming qubit 1 carries the distilled magic state)
\begin{equation*}
\begin{split}
&{\rho _{out}^{exp}} = \sum\limits_{i = 0}^7 {{\theta _i}} {\rho _i} \otimes \left| i \right\rangle \left\langle i \right| + \sum\limits_{i \ne j = 0}^7 {{\theta _{ij}}} {\rho _{ij}} \otimes \left| i \right\rangle \left\langle j \right|,\\
&{\rho _i} = \frac{1}{2}\left( {I + {x_{i}}{\sigma _x} + {z_{i}}{\sigma _z}} \right),
\end{split}
\end{equation*}
where ${{\theta _i}}$ is the probability of the measurement outcome, corresponding to the resulting state $\left| i \right\rangle $ of the other three redundant qubits, and $\left| i \right\rangle  = \left| {000} \right\rangle ,\left| {001} \right\rangle ,...,\left| {111} \right\rangle $, for $i = 0,1,2,...,7$. Measuring outcome $\left| {000} \right\rangle $ indicates a successful purification.
We can determine all the wanted parameters by following steps:

%\bigskip
\noindent \textbf{a.} Read out on each qubit after the application of a $\pi/2$ pulse;

By this step, we can get all diagonal elements ${\left\{ {{m_i}} \right\}_{i = 1,...,16}}$ of $\rho _{out}^{exp}$.

%\begin{table}[h]
%\includegraphics[scale=0.6]{tomo}
%\caption{\label{tomo} The integrated intensity of each peak in NMR spectra is proportional to the difference between the populations of two related energy
%levels. }
%\end{table}

For example, after operating a $\pi/2$ pulse on qubit 1, the intensity of the spectral line, which corresponds to the transition $\left| {0000} \right\rangle $ to $\left| {1000} \right\rangle $, is proportional to the difference between the corresponding populations, i.e. ${m_1} - {m_9}$.

Then we get all ${\theta _i} = {m_i} + {m_{i + 8}}$.

%\bigskip

\noindent \textbf{b.}	Read out on qubit 1 and read out on qubit 1 after the application of a $\pi/2$ pulse;
These two readout operations are sufficient to measure all ${\theta _i}{\rho _i}$.

\end{document}